\documentclass[twoside]{article}
\usepackage{epsfig,IFNA2012,graphicx,times,amsmath,amsthm, amssymb, multicol}
\def\ohead{Finite quantum mechanical model for the stock market}
\def\ehead{Liviu-Adrian Cotfas}
\def\received{August 31, 2012}

\begin{document}
\thispagestyle{myheadings}

%%%%%%%%%%%%%%%%%%%%%%%%%%%%%%
\setcounter{page}{1} %% You will be notified by conference coordinator of the actual page number
%%%%%%%%%%%%%%%%%%%%%%%%%%%%%%

\label{firstpage} %% Used in article header on first page

%\title{\ohead}
\title{Finite quantum mechanical\\ model for the stock market}
\author{Liviu-Adrian Cotfas\\ \\
Faculty of Economic Cybernetics, Statistics and Informatics, Academy of Economic Studies,\\ 
6 Piata Romana, 010374 Bucharest, Romania, E-mail:   \texttt{lcotfas@gmail.com} \\
}
\maketitle

\begin{abstract} 
The  price of a given stock is exactly known only at the time of sale when the stock is between the traders.  If we know the price (owner) then we have no information on the owner (price). A more general description including cases when we have partial information on both price and ownership is obtained by using the quantum mechanics methods. The relation price-ownership is similar to the relation position-momentum. Our approach is based on the mathematical formalism used in the case of quantum systems with finite-dimensional Hilbert space.  The linear operator corresponding to the ownership is obtained from the linear operator corresponding to the price by using the finite Fourier transform. In our idealized model, the Schr\" odinger type equation describing the time evolution of the stock price is solved numerically.

\vspace{2mm}
\end{abstract}

\pagestyle{headings}
\section{INTRODUCTION}

The mathematical modeling of price dynamics is a very complex problem. The information and market psychology \cite{Delcea} play important role in price dynamics, and therefore the methods of traditional financial dynamics are not sufficient for an adequate description. The extreme irregularities in the evolution of prices in financial markets can be better understood by using the mathematical formalism of quantum mechanics [2-7].  The time evolution of the price depends on many factors including the political environment, market information, economic policies  and psychology of traders. 

The presence of a  minimal trading value of stocks results in the possibility to use the discrete quantum mechanics methods \cite{Vourdas}. A given stock has not a definite price until it is traded. Its value is actualized as a consequence of the contextual interactions in the trading process \cite{Zhang,Cotfas1}. The usual description is in terms of probabilities, and the existence of probability interference suggests the necessity to use probability amplitudes, that is, wavefunctions.

\vspace{2mm}

\section{STOCK PRICE AND STOCK OWNERSHIP}

We consider a stock market with a large number $N$ of traders $T_0$, $T_1$,  ... ,$T_{N-1}$, and investigate the price of a fixed stock. 
By choosing an adequate unit of cash, we can assume that the only possible values of the price of the considered stock are $0$, $1$, $2$, ..., $N\!-\!1$. Following the analogy with the description of quantum systems \cite{Messiah}, we assume that the stock price at a fixed moment of time can be described by a function
\[
\Psi  :\{ 0,1,2,...,N\!-\!1\}\longrightarrow \mathbb{C}
\]
chosen such that $|\Psi (n)|^2$ represent the probability to have a price equal to $n$ units of cash if a transaction takes place. In order to have a consistent probabilistic interpretation, the function $\Psi $ must satisfy the relation
\[
\sum_{n=0}^{N-1}|\Psi (n)|^2=1
\]
that is, to be a normalized function. The finite Fourier transform  \cite{Vourdas,Mehta} allows us to associate to $\Psi $ the function
\[
F[\Psi ] :\{ 0,1,2,...,N\!-\!1\}\longrightarrow \mathbb{C}
\]
defined by the relation
\[
F[\Psi ](k)=\frac{1}{\sqrt{N}}\sum _{n=0}^{N-1}{\rm e}^{-\frac{2\pi {\rm i}}{N}nk}\, \Psi (n).
\]
The function $F[\Psi ]$ is normalized, and following the analogy with the description of quantum systems, we assume that  $|F[\Psi ](k)|^2$  represents the probability that the stock owner is $T_k$.

For example, in the case when the stock price is described by the function
\[
\Psi _1 :\{0,1,..., 20\}\longrightarrow \mathbb{C}, \qquad \Psi_1(n)=\left\{
\begin{array}{lll}
1 & {\rm for} & n=7\\
0 & {\rm for} & n\neq 7
\end{array} \right.
\]
the corresponding Fourier transform is
\[
\mathcal{F}[\Psi _1]:\{0,1,..., 20\}\longrightarrow \mathbb{C}, \qquad \mathcal{F}[\Psi _1](k)=\frac{1}{\sqrt{21}}\, {\rm e}^{-\frac{2\pi {\rm i}}{3}k}
\]
and we have (see Figure 1)
\[
|\Psi _1(n)|^2=\left\{
\begin{array}{lll}
1 & {\rm for} & n=7\\
0 & {\rm for} & n\neq 7
\end{array} \right.\qquad {\rm and}\qquad |\mathcal{F}[\Psi _1](k)|^2=\frac{1}{21}.
\]
This means that the stock price is $7$ (units of cash), and the probability for each trader to be the owner is the same, namely, $1/21$. We know with precision the price but we have no information on the stock owner.

\begin{center}
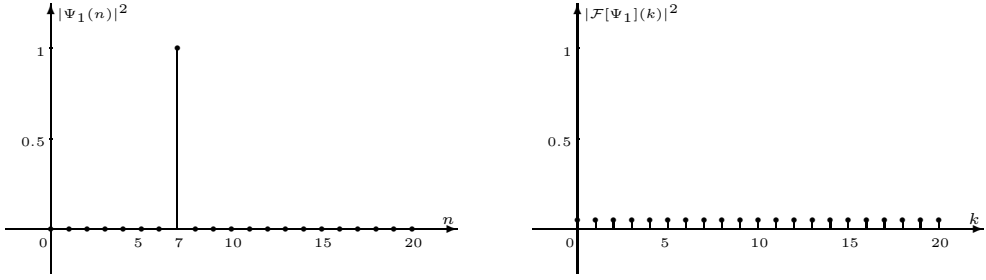
\begin{figure}[h]
\setlength{\unitlength}{2mm}
\begin{picture}(70,20)(0,0)
\put(2,3){\vector(1,0){30}}
\put(37,3){\vector(1,0){30}}

\put(5,0){\vector(0,1){18}}
\put(40,0){\vector(0,1){18}}

\put(    5.00000,   3){\circle*{0.3}}
\put(    6.20000,   3){\circle*{0.3}}
\put(    7.40000,   3){\circle*{0.3}}
\put(    8.60000,   3){\circle*{0.3}}
\put(    9.80000,   3){\circle*{0.3}}
\put(   11.00000,   3){\circle*{0.3}}
\put(   12.20000,  3){\circle*{0.3}}
\put(   13.40000,  15){\circle*{0.3}}
\put(   14.60000,   3){\circle*{0.3}}
\put(   15.80000,   3){\circle*{0.3}}
\put(   17.00000,   3){\circle*{0.3}}
\put(   18.20000,   3){\circle*{0.3}}
\put(   19.40000,   3){\circle*{0.3}}
\put(   20.60000,   3){\circle*{0.3}}
\put(   21.80000,   3){\circle*{0.3}}
\put(   23.00000,   3){\circle*{0.3}}
\put(   24.20000,   3){\circle*{0.3}}
\put(   25.40000,   3.00000){\circle*{0.3}}
\put(   26.60000,   3.00000){\circle*{0.3}}
\put(   27.80000,   3.00000){\circle*{0.3}}
\put(   29.00000,   3.00000){\circle*{0.3}}

\put(   13.40000,3){\line(0,1){   12}}

\put(   40.00000,   3.57){\circle*{0.3}}
\put(   41.20000,   3.57){\circle*{0.3}}
\put(   42.40000,   3.57){\circle*{0.3}}
\put(   43.60000,   3.57){\circle*{0.3}}
\put(   44.80000,   3.57){\circle*{0.3}}
\put(   46.00000,   3.57){\circle*{0.3}}
\put(   47.20000,   3.57){\circle*{0.3}}
\put(   48.40000,   3.57){\circle*{0.3}}
\put(   49.60000,   3.57){\circle*{0.3}}
\put(   50.80000,   3.57){\circle*{0.3}}
\put(   52.00000,   3.57){\circle*{0.3}}
\put(   53.20000,   3.57){\circle*{0.3}}
\put(   54.40000,   3.57){\circle*{0.3}}
\put(   55.60000,  3.57){\circle*{0.3}}
\put(   56.80000,  3.57){\circle*{0.3}}
\put(   58.00000,  3.57){\circle*{0.3}}
\put(   59.20000,   3.57){\circle*{0.3}}
\put(   60.40000,   3.57){\circle*{0.3}}
\put(   61.60000,   3.57){\circle*{0.3}}
\put(   62.80000,   3.57){\circle*{0.3}}
\put(   64.00000,   3.57){\circle*{0.3}}
\put(   40.00000,3){\line(0,1){    0.57}}
\put(   41.20000,3){\line(0,1){   0.57}}
\put(   42.40000,3){\line(0,1){    0.57}}
\put(   43.60000,3){\line(0,1){   0.57}}
\put(   44.80000,3){\line(0,1){   0.57}}
\put(   46.00000,3){\line(0,1){   0.57}}
\put(   47.20000,3){\line(0,1){   0.57}}
\put(   48.40000,3){\line(0,1){    0.57}}
\put(   49.60000,3){\line(0,1){    0.57}}
\put(   50.80000,3){\line(0,1){   0.57}}
\put(   52.00000,3){\line(0,1){   0.57}}
\put(   53.20000,3){\line(0,1){   0.57}}
\put(   54.40000,3){\line(0,1){   0.57}}
\put(   55.60000,3){\line(0,1){   0.57}}
\put(   56.80000,3){\line(0,1){  0.57}}
\put(   58.00000,3){\line(0,1){   0.57}}
\put(   59.20000,3){\line(0,1){   0.57}}
\put(   60.40000,3){\line(0,1){    0.57}}
\put(   61.60000,3){\line(0,1){   0.57}}
\put(   62.80000,3){\line(0,1){   0.57}}
\put(   64.00000,3){\line(0,1){   0.57}}

\put(4.2,1.8){$\scriptscriptstyle{0}$}
\put(28.5,1.8){$\scriptscriptstyle{20}$}
\put(16.5,1.8){$\scriptscriptstyle{10}$}
\put(10.5,1.8){$\scriptscriptstyle{5}$}
\put(13.2,1.8){$\scriptscriptstyle{7}$}
\put(22.5,1.8){$\scriptscriptstyle{15}$}

\put(39.2,1.8){$\scriptscriptstyle{0}$}
\put(63.5,1.8){$\scriptscriptstyle{20}$}
\put(51.5,1.8){$\scriptscriptstyle{10}$}
\put(45.5,1.8){$\scriptscriptstyle{5}$}
\put(57.5,1.8){$\scriptscriptstyle{15}$}

\put(31,3.3){$\scriptstyle{n}$}
\put(66,3.3){$\scriptstyle{k}$}

\put(3,8.5){$\scriptscriptstyle{0.5}$}
\put(4.9, 9){\line(1,0){0.2}}
\put(4,14.5){$\scriptscriptstyle{1}$}
\put(4.9, 15){\line(1,0){0.2}}
\put(38,8.5){$\scriptscriptstyle{0.5}$}
\put(39.9, 9){\line(1,0){0.2}}
\put(39,14.5){$\scriptscriptstyle{1}$}
\put(39.9, 15){\line(1,0){0.2}}

\put(5.4,17){$\scriptscriptstyle{|\Psi _1(n)|^2}$}
\put(40.4,17){$\scriptscriptstyle{|\mathcal{F}[\Psi _1](k)|^2}$}

\end{picture}
\caption{The distributions of probability $|\Psi _1(n)|^2$ and $|\mathcal{F}[\Psi _1](k)|^2$.}
\end{figure}
\end{center}

In the case when the stock price is described by  the function 
\[
\Psi _2  :\{0,1,..., 20\}\longrightarrow \mathbb{C}, \qquad \Psi_2(n)=\left\{
\begin{array}{cll}
\frac{1}{\sqrt{2}} & {\rm for} & n\!=\!7\\[1mm]
\frac{1}{2} & {\rm for} & n\!\in \!\{6,8\}\\[1mm]
0 & {\rm for} & n\!\not\in \!\{6,7,8\}
\end{array} \right.
\]
we have only partial information on the price
\[
|\Psi_2(n)|^2=\left\{
\begin{array}{lll}
\frac{1}{2} & {\rm for} & n\!=\!7\\[1mm]
\frac{1}{4} & {\rm for} & n\!\in \!\{6,8\}\\[1mm]
0 & {\rm for} & n\!\not\in \!\{6,7,8\}
\end{array} \right.
\]
but we have at the same time partial information on the owner.  The stock owner is $T_0$, $T_1$,  ... ,$T_{N-1}$ with the probabilities $|\mathcal{F}[\Psi _2](0)|^2$,  $|\mathcal{F}[\Psi _2](1)|^2$, ... , $|\mathcal{F}[\Psi _2](N\!-\!1)|^2$, respectively (see Figure 2). If a transaction  takes place, the stock price is $6$, $7$ or $8$ with the probabilities $\frac{1}{4}$, \ $\frac{1}{2}$ and $\frac{1}{4}$, respectively.

\begin{center}
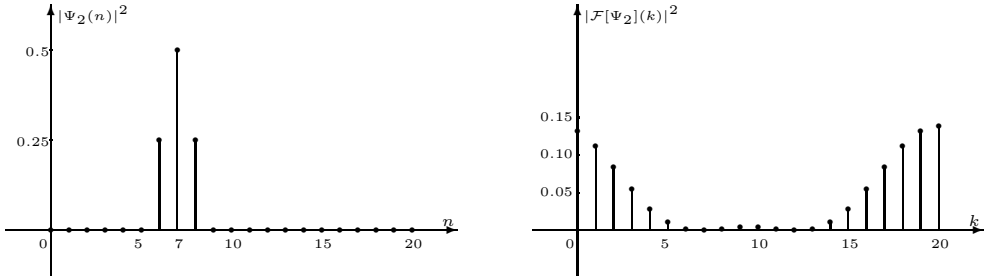
\begin{figure}[h]
\setlength{\unitlength}{2mm}
\begin{picture}(70,20)(0,0)
\put(2,3){\vector(1,0){30}}
\put(37,3){\vector(1,0){30}}

\put(5,0){\vector(0,1){18}}
\put(40,0){\vector(0,1){18}}

\put(    5.00000,   3){\circle*{0.3}}
\put(    6.20000,   3){\circle*{0.3}}
\put(    7.40000,   3){\circle*{0.3}}
\put(    8.60000,   3){\circle*{0.3}}
\put(    9.80000,   3){\circle*{0.3}}
\put(   11.00000,   3){\circle*{0.3}}
\put(   12.20000,  9){\circle*{0.3}}
\put(   13.40000,  15){\circle*{0.3}}
\put(   14.60000,   9){\circle*{0.3}}
\put(   15.80000,   3){\circle*{0.3}}
\put(   17.00000,   3){\circle*{0.3}}
\put(   18.20000,   3){\circle*{0.3}}
\put(   19.40000,   3){\circle*{0.3}}
\put(   20.60000,   3){\circle*{0.3}}
\put(   21.80000,   3){\circle*{0.3}}
\put(   23.00000,   3){\circle*{0.3}}
\put(   24.20000,   3){\circle*{0.3}}
\put(   25.40000,   3.00000){\circle*{0.3}}
\put(   26.60000,   3.00000){\circle*{0.3}}
\put(   27.80000,   3.00000){\circle*{0.3}}
\put(   29.00000,   3.00000){\circle*{0.3}}
\put(   12.20000,  3){\line(0,1){   6}}
\put(   13.40000,3){\line(0,1){   12}}
\put(   14.60000,   3){\line(0,1){   6}}
\put(   40.00000,   9.58215){\circle*{0.3}}
\put(   41.20000,   8.59795){\circle*{0.3}}
\put(   42.40000,   7.21545){\circle*{0.3}}
\put(   43.60000,   5.73844){\circle*{0.3}}
\put(   44.80000,   4.45540){\circle*{0.3}}
\put(   46.00000,   3.55910){\circle*{0.3}}
\put(   47.20000,   3.10213){\circle*{0.3}}
\put(   48.40000,   3.00160){\circle*{0.3}}
\put(   49.60000,   3.08948){\circle*{0.3}}
\put(   50.80000,   3.18897){\circle*{0.3}}
\put(   52.00000,   3.18897){\circle*{0.3}}
\put(   53.20000,   3.08948){\circle*{0.3}}
\put(   54.40000,   3.00160){\circle*{0.3}}
\put(   55.60000,   3.10213){\circle*{0.3}}
\put(   56.80000,   3.55910){\circle*{0.3}}
\put(   58.00000,   4.45540){\circle*{0.3}}
\put(   59.20000,   5.73844){\circle*{0.3}}
\put(   60.40000,   7.21545){\circle*{0.3}}
\put(   61.60000,   8.59795){\circle*{0.3}}
\put(   62.80000,   9.58215){\circle*{0.3}}
\put(   64.00000,   9.93860){\circle*{0.3}}
\put(   40.00000,3){\line(0,1){   6.58215}}
\put(   41.20000,3){\line(0,1){   5.59795}}
\put(   42.40000,3){\line(0,1){   4.21545}}
\put(   43.60000,3){\line(0,1){   2.73844}}
\put(   44.80000,3){\line(0,1){   1.45540}}
\put(   46.00000,3){\line(0,1){    .55910}}
\put(   47.20000,3){\line(0,1){    .10213}}
\put(   48.40000,3){\line(0,1){    .00160}}
\put(   49.60000,3){\line(0,1){    .08948}}
\put(   50.80000,3){\line(0,1){    .18897}}
\put(   52.00000,3){\line(0,1){    .18897}}
\put(   53.20000,3){\line(0,1){    .08948}}
\put(   54.40000,3){\line(0,1){    .00160}}
\put(   55.60000,3){\line(0,1){    .10213}}
\put(   56.80000,3){\line(0,1){    .55910}}
\put(   58.00000,3){\line(0,1){   1.45540}}
\put(   59.20000,3){\line(0,1){   2.73844}}
\put(   60.40000,3){\line(0,1){   4.21545}}
\put(   61.60000,3){\line(0,1){   5.59795}}
\put(   62.80000,3){\line(0,1){   6.58215}}
\put(   64.00000,3){\line(0,1){   6.93860}}

\put(4.2,1.8){$\scriptscriptstyle{0}$}
\put(28.5,1.8){$\scriptscriptstyle{20}$}
\put(16.5,1.8){$\scriptscriptstyle{10}$}
\put(10.5,1.8){$\scriptscriptstyle{5}$}
\put(13.2,1.8){$\scriptscriptstyle{7}$}
\put(22.5,1.8){$\scriptscriptstyle{15}$}

\put(39.2,1.8){$\scriptscriptstyle{0}$}
\put(63.5,1.8){$\scriptscriptstyle{20}$}
\put(51.5,1.8){$\scriptscriptstyle{10}$}
\put(45.5,1.8){$\scriptscriptstyle{5}$}
\put(57.5,1.8){$\scriptscriptstyle{15}$}

\put(31,3.3){$\scriptstyle{n}$}
\put(66,3.3){$\scriptstyle{k}$}

\put(2.7,8.6){$\scriptscriptstyle{0.25}$}
\put(4.9, 9){\line(1,0){0.2}}
\put(3.3,14.6){$\scriptscriptstyle{0.5}$}
\put(4.9, 15){\line(1,0){0.2}}
\put(37.5,10.3){$\scriptscriptstyle{0.15}$}
\put(39.9, 10.5){\line(1,0){0.2}}
\put(37.5,7.8){$\scriptscriptstyle{0.10}$}
\put(39.9, 8){\line(1,0){0.2}}
\put(37.5,5.3){$\scriptscriptstyle{0.05}$}
\put(39.9, 5.5){\line(1,0){0.2}}

\put(5.4,17){$\scriptscriptstyle{|\Psi _2(n)|^2}$}
\put(40.4,17){$\scriptscriptstyle{|\mathcal{F}[\Psi _2](k)|^2}$}

\end{picture}
\caption{The distributions of probability $|\Psi _2(n)|^2$ and $|\mathcal{F}[\Psi _2](k)|^2$.}
\end{figure}
\end{center}

The scalar product of two functions
\[
\Psi  :\{ 0,1,2,...,N\!-\!1\}\longrightarrow \mathbb{C}\qquad {\rm and}\qquad \Phi  :\{ 0,1,2,...,N\!-\!1\}\longrightarrow \mathbb{C}
\]
is defined by the formula
\[
\langle \Psi ,\Phi \rangle =\sum_{n=0}^{N-1}\overline{\Psi (n)}\, \Phi (n)
\]
where $\overline{\Psi (n)}$ is the complex conjugate of $\Psi (n)$. The space $\mathcal{H}$ of all the functions
\[
\Psi  :\{ 0,1,2,...,N\!-\!1\}\longrightarrow \mathbb{C}
\]
is a $N$-dimensional Hilbert space.

\vspace{2mm}

\section{PRICE AND OWNERSHIP OPERATORS}

In quantum mechanics, in the case of a particle moving along an axis, the position is described by the linear operator \cite{Messiah}
\[
 \psi \mapsto \hat x \, \psi\qquad  {\rm where}\qquad  (\hat x \, \psi )(x)=x\, \psi (x)
\]
and the momentum by the linear operator ($\hbar $ is Planck's constant $h$ divided by $2\pi $)
\[
\psi \mapsto \hat p \, \psi \qquad  {\rm where}\qquad  \hat p \, \psi =-{\rm i}\hbar \frac{d\psi }{dx}.
\]
In a state described by the  normalized wavefunction $\psi $, the numbers 
\[
\langle \hat x\rangle =\langle \psi ,\hat x\, \psi \rangle \qquad {\rm and} \qquad 
\langle \hat p\rangle =\langle \psi ,\hat p\, \psi \rangle 
\]
represent the mean value of coordinate and momentum, respectively.
The position and momentum operators  are related through the Fourier transform as
\[
\hat p=F^{-1}\hat xF.
\]
The relation price-ownership is, in a certain sense,  similar to the relation position-momentum: if we know the stock owner then we have no information on the price, and conversely, if we know the price we have no information on the stock owner. Nevertheless, we have to recognize that the numerical description we use for the ownership is rather artificial.

By following the analogy with quantum mechanics we define the {\em price operator} as
\[
\wp :\mathcal{H}\longrightarrow \mathcal{H}:\varphi \mapsto \wp\, \varphi ,\qquad (\wp\, \varphi )(n)=n\, \varphi (n)
\]
and the {\em ownership operator} as
\[
\mathcal{O} :\mathcal{H}\longrightarrow \mathcal{H}, \qquad \mathcal{O}=\mathcal{F}^{-1}\wp \mathcal{F}.
\]
If $\Phi $ is a normalized function then the number
\[
\langle \wp \rangle =\langle \Phi ,\wp \Phi \rangle =\sum _{n=0}^{N-1}n\, |\Phi (n)|^2
\]
represents the {\it mean value} of the  price.

\section{TIME EVOLUTION OF THE STOCK PRICE}

The main purpose of a mathematical model concerning the stock price is to anticipate the price.  We can only predict probabilities. By following the analogy with the quantum mechanics, we assume that the time dependent function $\Phi (n,t)$ describing the stock price satisfies a Schr\" odinger type equation \cite{Messiah}
\[
{\rm i}\, \frac{\partial }{\partial t}\Phi =\left( \frac{1}{2\mu } {\mathcal{O}}^2+\mathcal{V}(\wp ,t)\right)\Phi 
\]
where $\mu $ is a positive parameter and $\mathcal{V}(\wp ,t)$ is a time dependent function describing the interactions between traders as well as the external economic conditions. The function $\Phi (n,t)$ is well-determined if its values $\Phi (n,t_0)$ at a fixed moment of time $t_0$ are known.

A large variety of cases can be easily investigated by using, for example, the computer program in MATHEMATICA presented in \cite{Cotfas2}. The particular case of the equation
\[
{\rm i}\, \frac{\partial \Phi }{\partial t}=\left(  \frac{1}{2\mu } {\mathcal{O}}^2+\beta \, \cos \omega t\,\, \wp \right) \Phi 
\]
considered for $\alpha \!=\!0.2$, $\mu \!=\!1$, $\beta \!=\!1/10$ and  $\omega \!=\!1/10000$,
is ilustrated in Figure 3. At the initial moment of time, $t=0$, the price of the considered stock is 7 units of cash, but we do not have any information on the owner. After 4 hours the most probable price is 8, but the price may also be 7 or 9 with certain probabilities. The most probable owner is $T_{20}$ , but $T_0$, $T_1$, $T_2$, $T_{17}$, $T_{18}$, $T_{19}$ may also be the owners with smaller probabilities. After 8 hours the most probable price is 9 and the probability for $T_{20}$ to be the owner is 0.43.

\begin{center}
\begin{figure}[t]
\setlength{\unitlength}{1.8mm}
\begin{picture}(70,40)(-9,0)
\put(0,2){\vector(1,0){20}}
\put(23,2){\vector(1,0){20}}
\put(46,2){\vector(1,0){20}}
\put(1.5,0){\vector(0,1){17}}
\put(24.5,0){\vector(0,1){17}}
\put(47.5,0){\vector(0,1){17}}

\put(0,22){\vector(1,0){20}}
\put(23,22){\vector(1,0){20}}
\put(46,22){\vector(1,0){20}}
\put(1.5,20){\vector(0,1){17}}
\put(24.5,20){\vector(0,1){17}}
\put(47.5,20){\vector(0,1){17}}

\put(    1.50000,  22.00000){\circle*{0.3}}
\put(    2.30000,  22.00000){\circle*{0.3}}
\put(    3.10000,  22.00000){\circle*{0.3}}
\put(    3.90000,  22.00000){\circle*{0.3}}
\put(    4.70000,  22.00000){\circle*{0.3}}
\put(    5.50000,  22.00000){\circle*{0.3}}
\put(    6.30000,  22.00000){\circle*{0.3}}
\put(    7.10000,  34.00000){\circle*{0.3}}
\put(    7.90000,  22.00000){\circle*{0.3}}
\put(    8.70000,  22.00000){\circle*{0.3}}
\put(    9.50000,  22.00000){\circle*{0.3}}
\put(   10.30000,  22.00000){\circle*{0.3}}
\put(   11.10000,  22.00000){\circle*{0.3}}
\put(   11.90000,  22.00000){\circle*{0.3}}
\put(   12.70000,  22.00000){\circle*{0.3}}
\put(   13.50000,  22.00000){\circle*{0.3}}
\put(   14.30000,  22.00000){\circle*{0.3}}
\put(   15.10000,  22.00000){\circle*{0.3}}
\put(   15.90000,  22.00000){\circle*{0.3}}
\put(   16.70000,  22.00000){\circle*{0.3}}
\put(   17.50000,  22.00000){\circle*{0.3}}
\put(    1.50000,22){\line(0,1){    .00000}}
\put(    2.30000,22){\line(0,1){    .00000}}
\put(    3.10000,22){\line(0,1){    .00000}}
\put(    3.90000,22){\line(0,1){    .00000}}
\put(    4.70000,22){\line(0,1){    .00000}}
\put(    5.50000,22){\line(0,1){    .00000}}
\put(    6.30000,22){\line(0,1){    .00000}}
\put(    7.10000,22){\line(0,1){  12.00000}}
\put(    7.90000,22){\line(0,1){    .00000}}
\put(    8.70000,22){\line(0,1){    .00000}}
\put(    9.50000,22){\line(0,1){    .00000}}
\put(   10.30000,22){\line(0,1){    .00000}}
\put(   11.10000,22){\line(0,1){    .00000}}
\put(   11.90000,22){\line(0,1){    .00000}}
\put(   12.70000,22){\line(0,1){    .00000}}
\put(   13.50000,22){\line(0,1){    .00000}}
\put(   14.30000,22){\line(0,1){    .00000}}
\put(   15.10000,22){\line(0,1){    .00000}}
\put(   15.90000,22){\line(0,1){    .00000}}
\put(   16.70000,22){\line(0,1){    .00000}}
\put(   17.50000,22){\line(0,1){    .00000}}
\put(   24.50000,  22.20635){\circle*{0.3}}
\put(   25.30000,  22.10983){\circle*{0.3}}
\put(   26.10000,  22.07620){\circle*{0.3}}
\put(   26.90000,  22.06694){\circle*{0.3}}
\put(   27.70000,  22.07098){\circle*{0.3}}
\put(   28.50000,  22.09579){\circle*{0.3}}
\put(   29.30000,  22.20093){\circle*{0.3}}
\put(   30.10000,  22.98023){\circle*{0.3}}
\put(   30.90000,  31.28398){\circle*{0.3}}
\put(   31.70000,  22.75877){\circle*{0.3}}
\put(   32.50000,  22.08032){\circle*{0.3}}
\put(   33.30000,  22.01711){\circle*{0.3}}
\put(   34.10000,  22.00504){\circle*{0.3}}
\put(   34.90000,  22.00155){\circle*{0.3}}
\put(   35.70000,  22.00041){\circle*{0.3}}
\put(   36.50000,  22.00019){\circle*{0.3}}
\put(   37.30000,  22.00058){\circle*{0.3}}
\put(   38.10000,  22.00166){\circle*{0.3}}
\put(   38.90000,  22.00403){\circle*{0.3}}
\put(   39.70000,  22.00973){\circle*{0.3}}
\put(   40.50000,  22.02937){\circle*{0.3}}
\put(   24.50000,22){\line(0,1){    .20635}}
\put(   25.30000,22){\line(0,1){    .10983}}
\put(   26.10000,22){\line(0,1){    .07620}}
\put(   26.90000,22){\line(0,1){    .06694}}
\put(   27.70000,22){\line(0,1){    .07098}}
\put(   28.50000,22){\line(0,1){    .09579}}
\put(   29.30000,22){\line(0,1){    .20093}}
\put(   30.10000,22){\line(0,1){    .98023}}
\put(   30.90000,22){\line(0,1){   9.28398}}
\put(   31.70000,22){\line(0,1){    .75877}}
\put(   32.50000,22){\line(0,1){    .08032}}
\put(   33.30000,22){\line(0,1){    .01711}}
\put(   34.10000,22){\line(0,1){    .00504}}
\put(   34.90000,22){\line(0,1){    .00155}}
\put(   35.70000,22){\line(0,1){    .00041}}
\put(   36.50000,22){\line(0,1){    .00019}}
\put(   37.30000,22){\line(0,1){    .00058}}
\put(   38.10000,22){\line(0,1){    .00166}}
\put(   38.90000,22){\line(0,1){    .00403}}
\put(   39.70000,22){\line(0,1){    .00973}}
\put(   40.50000,22){\line(0,1){    .02937}}
\put(   47.50000,  22.67679){\circle*{0.3}}
\put(   48.30000,  22.10182){\circle*{0.3}}
\put(   49.10000,  22.00436){\circle*{0.3}}
\put(   49.90000,  22.03846){\circle*{0.3}}
\put(   50.70000,  22.08043){\circle*{0.3}}
\put(   51.50000,  22.14774){\circle*{0.3}}
\put(   52.30000,  22.30626){\circle*{0.3}}
\put(   53.10000,  22.79503){\circle*{0.3}}
\put(   53.90000,  24.22334){\circle*{0.3}}
\put(   54.70000,  27.44153){\circle*{0.3}}
\put(   55.50000,  23.71970){\circle*{0.3}}
\put(   56.30000,  22.27576){\circle*{0.3}}
\put(   57.10000,  22.05208){\circle*{0.3}}
\put(   57.90000,  22.01484){\circle*{0.3}}
\put(   58.70000,  22.00802){\circle*{0.3}}
\put(   59.50000,  22.00779){\circle*{0.3}}
\put(   60.30000,  22.00926){\circle*{0.3}}
\put(   61.10000,  22.01131){\circle*{0.3}}
\put(   61.90000,  22.01410){\circle*{0.3}}
\put(   62.70000,  22.02010){\circle*{0.3}}
\put(   63.50000,  22.05127){\circle*{0.3}}
\put(   47.50000,22){\line(0,1){    .67679}}
\put(   48.30000,22){\line(0,1){    .10182}}
\put(   49.10000,22){\line(0,1){    .00436}}
\put(   49.90000,22){\line(0,1){    .03846}}
\put(   50.70000,22){\line(0,1){    .08043}}
\put(   51.50000,22){\line(0,1){    .14774}}
\put(   52.30000,22){\line(0,1){    .30626}}
\put(   53.10000,22){\line(0,1){    .79503}}
\put(   53.90000,22){\line(0,1){   2.22334}}
\put(   54.70000,22){\line(0,1){   5.44153}}
\put(   55.50000,22){\line(0,1){   1.71970}}
\put(   56.30000,22){\line(0,1){    .27576}}
\put(   57.10000,22){\line(0,1){    .05208}}
\put(   57.90000,22){\line(0,1){    .01484}}
\put(   58.70000,22){\line(0,1){    .00802}}
\put(   59.50000,22){\line(0,1){    .00779}}
\put(   60.30000,22){\line(0,1){    .00926}}
\put(   61.10000,22){\line(0,1){    .01131}}
\put(   61.90000,22){\line(0,1){    .01410}}
\put(   62.70000,22){\line(0,1){    .02010}}
\put(   63.50000,22){\line(0,1){    .05127}}
\put(    1.50000,   2.95238){\circle*{0.3}}
\put(    2.30000,   2.95238){\circle*{0.3}}
\put(    3.10000,   2.95238){\circle*{0.3}}
\put(    3.90000,   2.95238){\circle*{0.3}}
\put(    4.70000,   2.95238){\circle*{0.3}}
\put(    5.50000,   2.95238){\circle*{0.3}}
\put(    6.30000,   2.95238){\circle*{0.3}}
\put(    7.10000,   2.95238){\circle*{0.3}}
\put(    7.90000,   2.95238){\circle*{0.3}}
\put(    8.70000,   2.95238){\circle*{0.3}}
\put(    9.50000,   2.95238){\circle*{0.3}}
\put(   10.30000,   2.95238){\circle*{0.3}}
\put(   11.10000,   2.95238){\circle*{0.3}}
\put(   11.90000,   2.95238){\circle*{0.3}}
\put(   12.70000,   2.95238){\circle*{0.3}}
\put(   13.50000,   2.95238){\circle*{0.3}}
\put(   14.30000,   2.95238){\circle*{0.3}}
\put(   15.10000,   2.95238){\circle*{0.3}}
\put(   15.90000,   2.95238){\circle*{0.3}}
\put(   16.70000,   2.95238){\circle*{0.3}}
\put(   17.50000,   2.95238){\circle*{0.3}}
\put(    1.50000,2){\line(0,1){    .95238}}
\put(    2.30000,2){\line(0,1){    .95238}}
\put(    3.10000,2){\line(0,1){    .95238}}
\put(    3.90000,2){\line(0,1){    .95238}}
\put(    4.70000,2){\line(0,1){    .95238}}
\put(    5.50000,2){\line(0,1){    .95238}}
\put(    6.30000,2){\line(0,1){    .95238}}
\put(    7.10000,2){\line(0,1){    .95238}}
\put(    7.90000,2){\line(0,1){    .95238}}
\put(    8.70000,2){\line(0,1){    .95238}}
\put(    9.50000,2){\line(0,1){    .95238}}
\put(   10.30000,2){\line(0,1){    .95238}}
\put(   11.10000,2){\line(0,1){    .95238}}
\put(   11.90000,2){\line(0,1){    .95238}}
\put(   12.70000,2){\line(0,1){    .95238}}
\put(   13.50000,2){\line(0,1){    .95238}}
\put(   14.30000,2){\line(0,1){    .95238}}
\put(   15.10000,2){\line(0,1){    .95238}}
\put(   15.90000,2){\line(0,1){    .95238}}
\put(   16.70000,2){\line(0,1){    .95238}}
\put(   17.50000,2){\line(0,1){    .95238}}
\put(   24.50000,   4.10278){\circle*{0.3}}
\put(   25.30000,   3.34177){\circle*{0.3}}
\put(   26.10000,   3.22258){\circle*{0.3}}
\put(   26.90000,   2.59316){\circle*{0.3}}
\put(   27.70000,   2.60891){\circle*{0.3}}
\put(   28.50000,   2.41453){\circle*{0.3}}
\put(   29.30000,   2.28435){\circle*{0.3}}
\put(   30.10000,   2.32106){\circle*{0.3}}
\put(   30.90000,   2.19941){\circle*{0.3}}
\put(   31.70000,   2.23679){\circle*{0.3}}
\put(   32.50000,   2.23679){\circle*{0.3}}
\put(   33.30000,   2.19941){\circle*{0.3}}
\put(   34.10000,   2.32106){\circle*{0.3}}
\put(   34.90000,   2.28435){\circle*{0.3}}
\put(   35.70000,   2.41453){\circle*{0.3}}
\put(   36.50000,   2.60891){\circle*{0.3}}
\put(   37.30000,   2.59316){\circle*{0.3}}
\put(   38.10000,   3.22258){\circle*{0.3}}
\put(   38.90000,   3.34177){\circle*{0.3}}
\put(   39.70000,   4.10278){\circle*{0.3}}
\put(   40.50000,   7.34932){\circle*{0.3}}
\put(   24.50000,2){\line(0,1){   2.10278}}
\put(   25.30000,2){\line(0,1){   1.34177}}
\put(   26.10000,2){\line(0,1){   1.22258}}
\put(   26.90000,2){\line(0,1){    .59316}}
\put(   27.70000,2){\line(0,1){    .60891}}
\put(   28.50000,2){\line(0,1){    .41453}}
\put(   29.30000,2){\line(0,1){    .28435}}
\put(   30.10000,2){\line(0,1){    .32106}}
\put(   30.90000,2){\line(0,1){    .19941}}
\put(   31.70000,2){\line(0,1){    .23679}}
\put(   32.50000,2){\line(0,1){    .23679}}
\put(   33.30000,2){\line(0,1){    .19941}}
\put(   34.10000,2){\line(0,1){    .32106}}
\put(   34.90000,2){\line(0,1){    .28435}}
\put(   35.70000,2){\line(0,1){    .41453}}
\put(   36.50000,2){\line(0,1){    .60891}}
\put(   37.30000,2){\line(0,1){    .59316}}
\put(   38.10000,2){\line(0,1){   1.22258}}
\put(   38.90000,2){\line(0,1){   1.34177}}
\put(   39.70000,2){\line(0,1){   2.10278}}
\put(   40.50000,2){\line(0,1){   5.34932}}
\put(   47.50000,   4.33362){\circle*{0.3}}
\put(   48.30000,   3.89599){\circle*{0.3}}
\put(   49.10000,   2.63365){\circle*{0.3}}
\put(   49.90000,   2.29577){\circle*{0.3}}
\put(   50.70000,   2.31891){\circle*{0.3}}
\put(   51.50000,   2.01022){\circle*{0.3}}
\put(   52.30000,   2.14134){\circle*{0.3}}
\put(   53.10000,   2.01048){\circle*{0.3}}
\put(   53.90000,   2.04897){\circle*{0.3}}
\put(   54.70000,   2.02340){\circle*{0.3}}
\put(   55.50000,   2.02340){\circle*{0.3}}
\put(   56.30000,   2.04897){\circle*{0.3}}
\put(   57.10000,   2.01048){\circle*{0.3}}
\put(   57.90000,   2.14134){\circle*{0.3}}
\put(   58.70000,   2.01022){\circle*{0.3}}
\put(   59.50000,   2.31891){\circle*{0.3}}
\put(   60.30000,   2.29577){\circle*{0.3}}
\put(   61.10000,   2.63365){\circle*{0.3}}
\put(   61.90000,   3.89599){\circle*{0.3}}
\put(   62.70000,   4.33362){\circle*{0.3}}
\put(   63.50000,  10.57528){\circle*{0.3}}
\put(   47.50000,2){\line(0,1){   2.33362}}
\put(   48.30000,2){\line(0,1){   1.89599}}
\put(   49.10000,2){\line(0,1){    .63365}}
\put(   49.90000,2){\line(0,1){    .29577}}
\put(   50.70000,2){\line(0,1){    .31891}}
\put(   51.50000,2){\line(0,1){    .01022}}
\put(   52.30000,2){\line(0,1){    .14134}}
\put(   53.10000,2){\line(0,1){    .01048}}
\put(   53.90000,2){\line(0,1){    .04897}}
\put(   54.70000,2){\line(0,1){    .02340}}
\put(   55.50000,2){\line(0,1){    .02340}}
\put(   56.30000,2){\line(0,1){    .04897}}
\put(   57.10000,2){\line(0,1){    .01048}}
\put(   57.90000,2){\line(0,1){    .14134}}
\put(   58.70000,2){\line(0,1){    .01022}}
\put(   59.50000,2){\line(0,1){    .31891}}
\put(   60.30000,2){\line(0,1){    .29577}}
\put(   61.10000,2){\line(0,1){    .63365}}
\put(   61.90000,2){\line(0,1){   1.89599}}
\put(   62.70000,2){\line(0,1){   2.33362}}
\put(   63.50000,2){\line(0,1){   8.57528}}

%%%%%%%%%%%%%%%%%%%%%

\put(12,34){$\scriptscriptstyle{t=0}$}
\put(2,33.5){$\scriptscriptstyle{1}$}
\put( 1.3,34){\line(1,0){0.4}}
\put(2,27.5){$\scriptscriptstyle{0.5}$}
\put( 1.3,28){\line(1,0){0.4}}
\put(19,22.5){$\scriptstyle{n}$}
\put(2,36){$\scriptstyle{|\Psi(n,t)|^2}$}

\put(33,34){$\scriptscriptstyle{t=240\, {\rm minutes}}$}
\put(25,33.5){$\scriptscriptstyle{1}$}
\put( 24.3,34){\line(1,0){0.4}}
\put(25,27.5){$\scriptscriptstyle{0.5}$}
\put( 24.3,28){\line(1,0){0.4}}
\put(42,22.5){$\scriptstyle{n}$}
\put(25,36){$\scriptstyle{|\Psi(n,t)|^2}$}

\put(56,34){$\scriptscriptstyle{t=480\, {\rm minutes}}$}
\put(48,33.5){$\scriptscriptstyle{1}$}
\put( 47.3,34){\line(1,0){0.4}}
\put(48,27.5){$\scriptscriptstyle{0.5}$}
\put( 47.3,28){\line(1,0){0.4}}
\put(65,22.5){$\scriptstyle{n}$}
\put(48,36){$\scriptstyle{|\Psi(n,t)|^2}$}
%%%%%%%%%%%%%%%
\put(0.7,0.9){$\scriptscriptstyle{0}$}
\put(17,0.9){$\scriptscriptstyle{20}$}
\put(23.7,0.9){$\scriptscriptstyle{0}$}
\put(40,0.9){$\scriptscriptstyle{20}$}
\put(46.7,0.9){$\scriptscriptstyle{0}$}
\put(63,0.9){$\scriptscriptstyle{20}$}
\put(5.3,0.9){$\scriptscriptstyle{5}$}
\put(13.1,0.9){$\scriptscriptstyle{15}$}
\put(28.3,0.9){$\scriptscriptstyle{5}$}
\put(36,0.9){$\scriptscriptstyle{15}$}
\put(51.3,0.9){$\scriptscriptstyle{5}$}
\put(59.2,0.9){$\scriptscriptstyle{15}$}
\put(9,0.9){$\scriptscriptstyle{10}$}
\put(32,0.9){$\scriptscriptstyle{10}$}
\put(55,0.9){$\scriptscriptstyle{10}$}

\put(0.7,20.9){$\scriptscriptstyle{0}$}
\put(17,20.9){$\scriptscriptstyle{20}$}
\put(23.7,20.9){$\scriptscriptstyle{0}$}
\put(40,20.9){$\scriptscriptstyle{20}$}
\put(46.7,20.9){$\scriptscriptstyle{0}$}
\put(63,20.9){$\scriptscriptstyle{20}$}
\put(5.3,20.9){$\scriptscriptstyle{5}$}
\put(13.1,20.9){$\scriptscriptstyle{15}$}
\put(28.3,20.9){$\scriptscriptstyle{5}$}
\put(36,20.9){$\scriptscriptstyle{15}$}
\put(51.3,20.9){$\scriptscriptstyle{5}$}
\put(59.2,20.9){$\scriptscriptstyle{15}$}
\put(9,20.9){$\scriptscriptstyle{10}$}
\put(32,20.9){$\scriptscriptstyle{10}$}
\put(55,20.9){$\scriptscriptstyle{10}$}

\put(12,14){$\scriptscriptstyle{t=0}$}
\put(2,11.5){$\scriptscriptstyle{0.5}$}
\put( 1.3,12){\line(1,0){0.4}}
\put(2,9.5){$\scriptscriptstyle{0.4}$}
\put( 1.3,10){\line(1,0){0.4}}
\put(2,7.5){$\scriptscriptstyle{0.3}$}
\put( 1.3,8){\line(1,0){0.4}}
\put(2,5.5){$\scriptscriptstyle{0.2}$}
\put( 1.3,6){\line(1,0){0.4}}
\put(2,3.5){$\scriptscriptstyle{0.1}$}
\put( 1.3,4){\line(1,0){0.4}}
\put(19,2.5){$\scriptstyle{k}$}
\put(2,16){$\scriptstyle{|F[\Psi ](k,t)|^2}$}

\put(33,14){$\scriptscriptstyle{t=240\, {\rm minutes}}$}
\put(25,11.5){$\scriptscriptstyle{0.5}$}
\put( 24.3,12){\line(1,0){0.4}}
\put(25,9.5){$\scriptscriptstyle{0.4}$}
\put( 24.3,10){\line(1,0){0.4}}
\put(25,7.5){$\scriptscriptstyle{0.3}$}
\put( 24.3,8){\line(1,0){0.4}}
\put(25,5.5){$\scriptscriptstyle{0.2}$}
\put( 24.3,6){\line(1,0){0.4}}
\put(42,2.5){$\scriptstyle{k}$}
\put(25,16){$\scriptstyle{|F[\Psi ](k,t)|^2}$}

\put(56,14){$\scriptscriptstyle{t=480\, {\rm minutes}}$}
\put(48,11.5){$\scriptscriptstyle{0.5}$}
\put( 47.3,12){\line(1,0){0.4}}
\put(48,9.5){$\scriptscriptstyle{0.4}$}
\put( 47.3,10){\line(1,0){0.4}}
\put(48,7.5){$\scriptscriptstyle{0.3}$}
\put( 47.3,8){\line(1,0){0.4}}
\put(48,5.5){$\scriptscriptstyle{0.2}$}
\put( 47.3,6){\line(1,0){0.4}}
\put(65,2.5){$\scriptstyle{k}$}
\put(48,16){$\scriptstyle{|F[\Psi ](k,t)|^2}$}
\end{picture}
\caption{Time evolution of the distributions of probability describing the price and ownership.}
\end{figure}
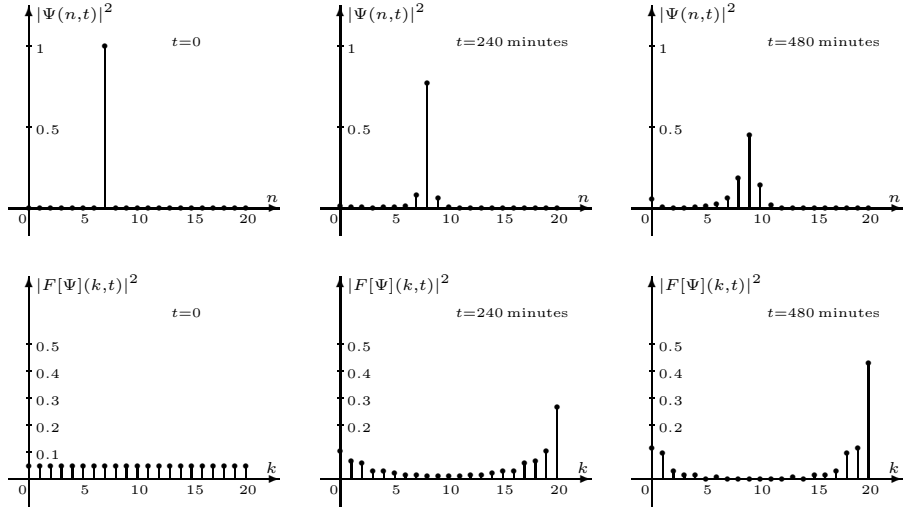
\end{center}

\vspace{2mm}

\section{CONCLUDING REMARKS}

The relation price-ownership is similar to the relation position-momentum from quantum mechanics. Despite the rather artificial numerical description we use for the ownership, the proposed quantum mechanical model may help us to better understand the relation price-ownership playing a fundamental role in finance. The distribution probability concerning the price is related to the distribution probability describing the ownership. The price and the ownership cannot be simultaneously known with infinite precision. Generaly, the accuracy in the knowledge of the price cannot be improved without a corresponding loss in the accuracy in the knowledge of the ownership.

%%%%%%%%%%%%%%%%%%%%%%%%%%%%%%%%%%%%%%%%%%%%%%%%%%%%%%%%%%%%%%%%%%%%%%%%%%%%%%%%%%%%%%%%%%%%%%%%%%%% 

\label{lastpage} 
\end{document}